\UseRawInputEncoding
\documentclass[reprent,aps,prl,twocolumn,
superscriptaddress,nofootinbib]{revtex4-2}
\usepackage{amsmath,amssymb}
\usepackage{graphicx}
\usepackage{bm}
\usepackage{hyperref}

\begin{document}

\title{Ferrochiral, antiferrochiral, and ferrichiral skyrmion crystals\\ in an itinerant  honeycomb magnet}

\author{Ryota Yambe}
\affiliation{Department of Applied Physics, The University of Tokyo, Tokyo 113-8656, Japan }
\author{Satoru Hayami}
\affiliation{Graduate School of Science, Hokkaido University, Sapporo 060-0810, Japan}

\begin{abstract}
Topological spin textures, such as a skyrmion crystal, are a source of unusual physical phenomena owing to the interplay between magnetism and topology.
Since physical phenomena depend on the topological property and the symmetry of underlying spin structures, the search for new topological spin textures and emergent phenomena is one of the challenges in condensed matter physics.
In this letter, we theoretically explore new topological spin textures arising from the synergy between spin, charge, and sublattice degrees of freedom in an itinerant magnet.
By performing simulated annealing for an effective spin model of the honeycomb Kondo lattice model, we find a plethora of skyrmion crystal instabilities at low temperatures, whose topological spin textures are classified into three types: ferrochiral, antiferrochiral, and ferrichiral skyrmion crystals. 
We show that the obtained skyrmion crystals are the consequence of the spin-orbit-coupling-free honeycomb structure.
Our results reveal the potential for itinerant honeycomb magnets to host a wide variety of SkXs and emergent phenomena.
\end{abstract}

\maketitle

\begin{figure}[t!]
\begin{center}
\includegraphics[width=1.0\hsize]{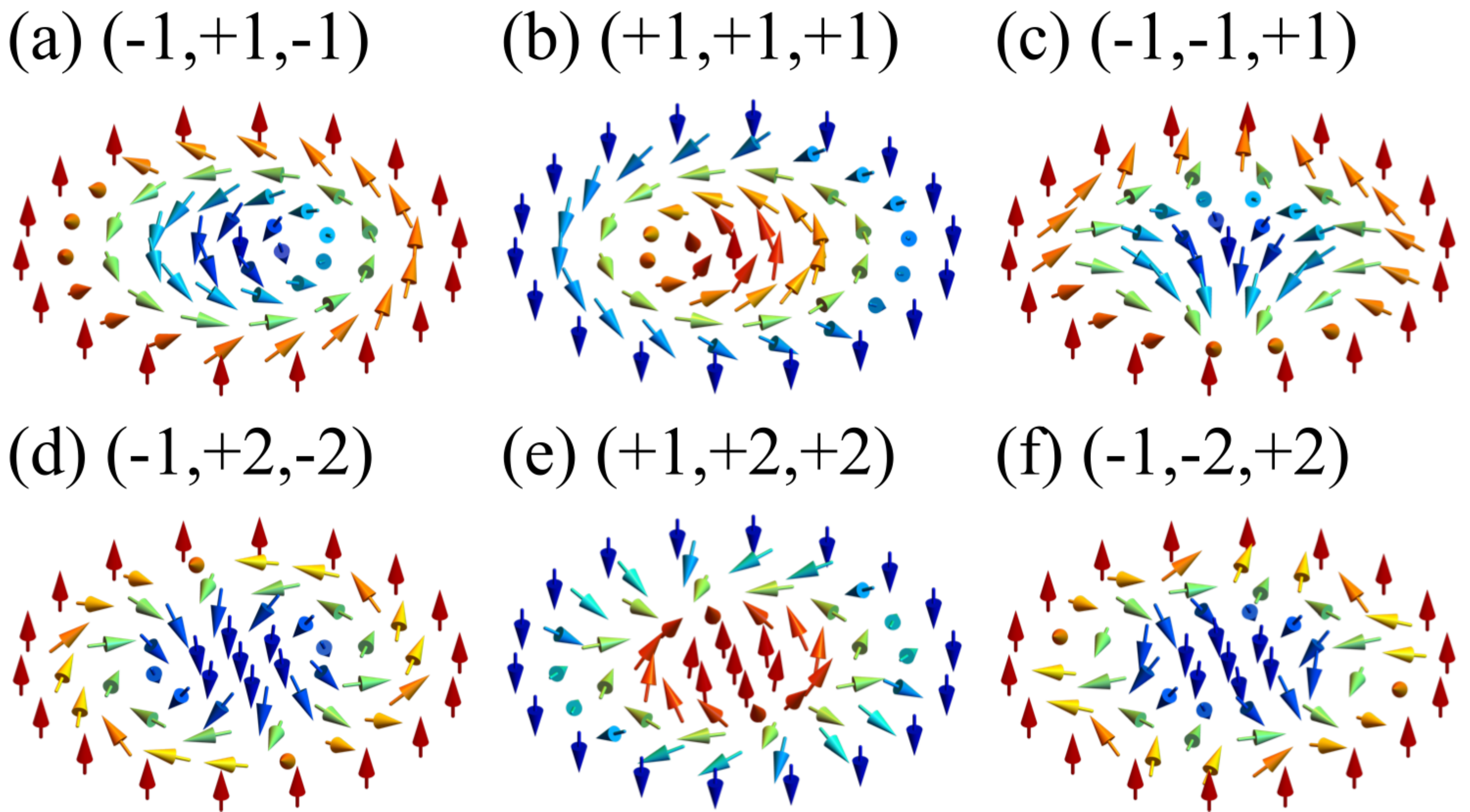} 
\caption{
\label{fig:skyrmion}
Skyrmions characterized by $(p,v,N_\mathrm{sk})$.
The arrow and color show the spin direction and $z$ component, respectively, where red (blue) corresponds to the positive (negative) $z$ component. 
}
\end{center}
\end{figure}

Noncoplanar spin structures with nontrivial topology (topological spin textures) have attracted much attention, since they give rise to fascinating physical phenomena arising from their emergent electromagnetic fields~\cite{nagaosa2013topological,Tokura_doi:10.1021/acs.chemrev.0c00297,zhang2020skyrmion,gobel2021beyond,hayami2021topological}.
The most familiar topological spin texture is a skyrmion, whose topological property is characterized by the skyrmion number $N_\mathrm{sk}=pv$; $p$ and $v$ represent the polarity and vorticity of the skyrmion, respectively~\cite{nagaosa2013topological}. 
Since the different sets of $(p,v,N_\mathrm{sk})$ result in different types of skyrmions, as shown in Figs.~\ref{fig:skyrmion}(a)-\ref{fig:skyrmion}(f), they become a source of  a variety of quantum transports and dynamics including the topological Hall/Nernst effect and skyrmion Hall effect~\cite{onoda2004anomalous,Neubauer_PhysRevLett.102.186602,Yi_PhysRevB.80.054416,Hamamoto_PhysRevB.92.115417,Shiomi_PhysRevB.88.064409,mizuta2016large,kim2018asymmetric,jin2019current,huang2017stabilization,weissenhofer2019orientation}.

The appearance of each skyrmion is dependent on a microscopic interaction. 
Specifically, the vorticity is determined by types of spin interactions. 
In centrosymmetric magnets, the competing exchange interactions in frustrated magnets lead to the crystal formation of the skyrmion (SkX) with $v=\pm 1$~\cite{Okubo_PhysRevLett.108.017206, leonov2015multiply, Lin_PhysRevB.93.064430, Hayami_PhysRevB.93.184413, Mitsumoto_PhysRevB.104.184432,Mitsumoto_PhysRevB.105.094427}, while the long-range higher-order exchange interaction in itinerant magnets favors the SkX with both $v=\pm 1$ and $\pm 2$ depending on the magnetic field~\cite{Ozawa_PhysRevLett.118.147205,Hayami_PhysRevB.95.224424,hayami2021topological}. 
The degeneracy in terms of $v$ is lifted by anisotropic exchange interactions originating from the spin-orbit coupling~\cite{Bogdanov89,Hayami_PhysRevLett.121.137202,Hayami_doi:10.7566/JPSJ.89.103702,yambe2022effective} or dipolar interactions~\cite{Utesov_PhysRevB.103.064414,Utesov_PhysRevB.105.054435}.  
Meanwhile, the polarity is determined by the magnetic field direction; $p=+1$ $(-1)$ is favored under the field along the $-z$ ($+z$) direction. 
In this way, a key essence to engineering $(p,v,N_\mathrm{sk})$ of the skyrmion has been clarified based on the microscopic interaction.

In this Letter, we explore a new type of SkXs by focusing on the sublattice degree of freedom. 
The effect of the sublattice degree of freedom has been studied in the 120$^\circ$ three-sublattice structure on the triangular lattice~\cite{Rosales_PhysRevB.92.214439}, honeycomb structure~\cite{Gobel_PhysRevB.96.060406,Shimokawa_PhysRevB.100.224404}, kagome structure~\cite{PhysRevB.100.245106}, diamond structure~\cite{Rosales_PhysRevB.105.224402}, and multi-layer structure~\cite{PhysRevB.104.014410,PhysRevX.12.031020,PhysRevB.105.224411,Hayami_PhysRevB.105.184426,PhysRevB.105.224411,hayami2022square,okigami2022engineering,lin2021skyrmion}.
One of the typical examples characteristic of the sublattice structure is the antiferromagnetic SkX on a bipartite lattice with the sublattice $\alpha=\mathrm{A}, \mathrm{B}$, where the sublattice A forms the SkX with $(p^{\rm A},v^{\rm A},N^{\rm A}_\mathrm{sk})=(-1,+1,-1)$ in Fig.~\ref{fig:skyrmion}(a) and the sublattice B forms that with $(p^{\rm B},v^{\rm B},N^{\rm B}_\mathrm{sk})=(+1,+1,+1)$ in Fig.~\ref{fig:skyrmion}(b). 
In this case, no topological Hall effect occurs due to the staggered skyrmion number $N^{\mathrm{A}}_\mathrm{sk}=-N^{\mathrm{B}}_\mathrm{sk}$, while the topological spin Hall effect is expected~\cite{Gobel_PhysRevB.96.060406}.

To realize a variety of topological spin textures in multi-sublattice systems, we focus on the synergy among the spin, charge, and sublattice degrees of freedom in an itinerant magnet.
By performing simulated annealing for an effective spin model of the honeycomb Kondo lattice model (KLM), we find that the synergy gives rise to five new SkXs with different $(N^\mathrm{A}_\mathrm{sk},N^\mathrm{B}_\mathrm{sk})$: two antiferrochiral SkXs with $(N^\mathrm{A}_\mathrm{sk},N^\mathrm{B}_\mathrm{sk})=(-1,+1)$ and $(-2,+2)$, two ferrochiral SkXs with $(N^\mathrm{A}_\mathrm{sk},N^\mathrm{B}_\mathrm{sk})=(-1,-1)$ and $(-2,-2)$, and a ferrichiral SkX with $(N^\mathrm{A}_\mathrm{sk},N^\mathrm{B}_\mathrm{sk})=(-2,+1)$.

\begin{figure}[t!]
\begin{center}
\includegraphics[width=1.0\hsize]{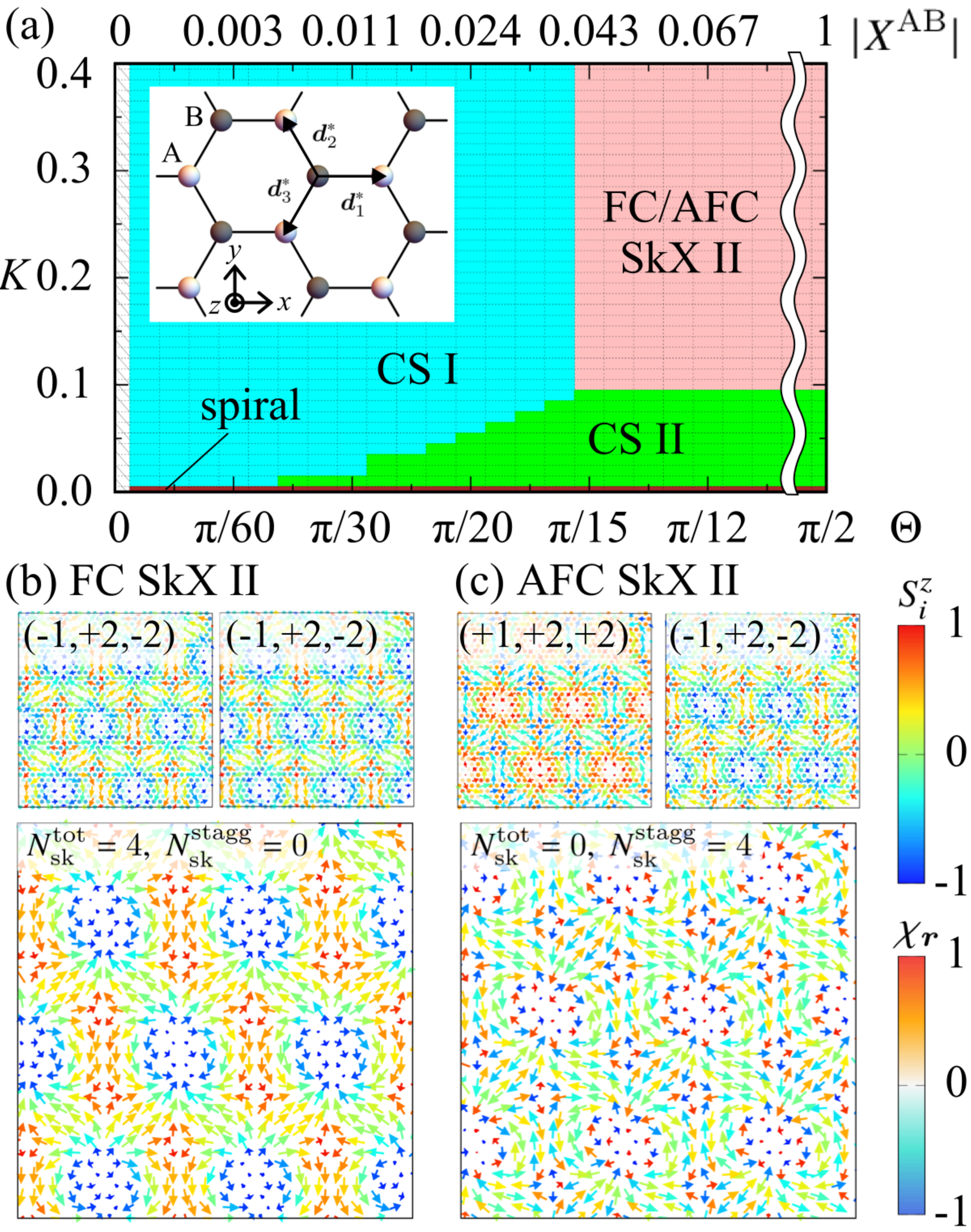} 
\caption{
\label{fig:Zero}
(a) Phase diagram on the $\Theta$-$K$ plane at $T=0.01$.
CS, FC SkX, and AFC SkX represent the chiral strip state, ferrochiral SkX, and antiferrochiral SkX, respectively. 
Inset shows the honeycomb struture with sublattices A (white) and B (gray).   
$\bm{d}^*_\eta$ ($\eta=1$-3) is the vector of three nearest-neighbor bonds.
Snapshots of (b) the FC and (c) AFC SkXs II at $\Theta=\pi/12$ and $K=0.4$. 
Upper left (right) panel: spin ($\bm{S}_i$) and chirality ($\chi_{\bm{r}}$) configurations on the sublattice A (B).
Lower panel: spin configuration on the honeycomb lattice.
The arrows, contours of arrows, and contours of circles show the $xy$ spin component, $z$ spin components, and spin scalar chirality, respectively.
$(p^{\alpha},v^{\alpha},N^{\alpha}_\mathrm{sk})$ and $(N^\mathrm{tot}_\mathrm{sk},N^\mathrm{stagg}_\mathrm{sk})$ are shown in the upper and lower panels, respectively.
}
\end{center}
\end{figure}

First, we consider the honeycomb lattice with the sublattices A and B shown in the inset of Fig.~\ref{fig:Zero}(a). The effective spin model, which is obtained in the weak-coupling regime of the honeycomb KLM, is given by 
\begin{align}
\label{eq:BBQM}
\mathcal{H}^\mathrm{eff}&=-2J\sum_{\eta}\Gamma_\eta (X)
+2\frac{K}{N}\sum_{\eta}\Gamma_\eta(X)^2,
\end{align} 
where $\Gamma_\eta(X)=\sum_{\alpha,\beta}X^{\alpha\beta}\bm{S}_{\alpha \bm{Q}_\eta}\cdot\bm{S}_{\beta -\bm{Q}_\eta}$; $\bm{S}_{\alpha \bm{q}}$ with the wave vector $\bm{q}$ and the sublattice $\alpha={\rm A}, {\rm B}$ is the Fourier transform of the localized spin, $X^{\alpha\beta}$ represents the form factor of the interaction in terms of the sublattice, $X^{{\rm AA}}=X^{{\rm BB}}$ and $X^{{\rm AB}}=(X^{{\rm BA}})^*$, and $N$ is the number of unit cells. 
The first term represents the bilinear interaction with $J>0$, which corresponds to the Ruderman-Kittel-Kasuya-Yosida (RKKY) interaction~\cite{Ruderman, Kasuya, Yosida1957}.
We only consider the dominant contributions at specific $\bm{Q}_\eta$, which gives the largest eigenvalue of the bare magnetic susceptibility of itinerant electrons in momentum space. 
We choose threefold symmetric ordering vectors so as to satisfy the honeycomb lattice symmetry: $\bm{Q}_1=(0,\pi/3)$, $\bm{Q}_2=(-\sqrt{3}\pi/6,-\pi/6)$, and $\bm{Q}_3=(\sqrt{3}\pi/6,-\pi/6)$, where we set the lattice constant as unity. 
Then, the form factor satisfies $X^{\alpha\beta}=(X^{\alpha\beta})^*$ due to the mirror symmetry with respect to the $xz$ plane~\cite{yambe2022effective}.
The second term represents the positive biquadratic interaction with $K>0$, which corresponds to the higher-order RKKY interaction and tends to favor noncoplanar spin textures including the SkX~\cite{Ozawa_PhysRevLett.118.147205,Hayami_PhysRevB.95.224424,hayami2021topological,Akagi_PhysRevLett.108.096401}.

In contrast to the previous effective spin model for the triangular KLM~\cite{Hayami_PhysRevB.95.224424}, the present model includes the effect of the intersublattice RKKY and biquadratic interactions owing to the multi-sublattice honeycomb structure.
Thus, the present model can describe the multiple-$Q$ instability that arises from the synergy between the spin, charge, and sublattice degrees of freedom.
In the following, we study the general case of the interactions by setting $X^{\rm AA}\equiv \cos^2\Theta$ and $X^{\rm AB}\equiv \pm \sin^2\Theta$ while changing $\Theta$ ($0\leq \Theta \leq \pi/2$) as well as $K$. 
We set $J=1$ as the energy unit. 
It is noted that the form factors in the first and second terms in Eq.~(\ref{eq:BBQM}) are usually different from each other, while the results are qualitatively similar even when considering different form factors (see Supplemental Material).

At $K=0$, the model in Eq.~(\ref{eq:BBQM}) exhibits the instability toward the single-$Q$ spiral state on each sublattice irrespective of $\Theta$. 
When $X^\mathrm{AB}\neq0$ ($\Theta>0$), both two spirals are characterized by the same $\bm{Q}_\eta$ and the same spiral plane for both positive and negative $X^\mathrm{AB}$. 
The sign dependence of $X^\mathrm{AB}$ is found in the relative spiral angle; the B spin at $\bm{R}_{\mathrm{B}i}$ is (anti)parallel to the A spin at $\bm{R}_{\mathrm{B}i}+\bm{d}^*_\eta$ for positive (negative) $X^\mathrm{AB}$, where $\bm{R}_{\mathrm{B}i}$ is the position vector at site $i$ on the sublattice B and $\bm{d}^*_\eta$ ($\bm{d}^*_\eta\cdot\bm{Q}_\eta=0$) is the displacement vector for three nearest-neighbor bonds shown in the inset of Fig.~\ref{fig:Zero}(a). 
In other words, positive (negative) $X^\mathrm{AB}$ tends to favor the (anti)ferromagnetic spin alignment for A and B sublattices.

Next, to investigate the ground state for nonzero $K$, we perform simulated annealing combined with the standard Metropolis local updates for the system size with $N=36^2$.
We gradually reduce the temperature with a rate $T_{n+1}=\alpha T_n$, where $T_n$ is the temperature at the $n$th step. 
We set the initial temperature $T_0=1$ and the coefficient $\alpha\approx0.999539589
$.
A final temperature $T_\mathrm{f} = 0.01$ is reached after total $10^6$ Monte Carlo steps (MCS), where we perform $10^2$ MCS at each temperature  $T_n$. 
After $10^5$ MCS for the thermalization at $T_\mathrm{f}$, we perform $10^6$ MCS for measurements, where $10^4$ samples are used for average.
To determine the phase boundary, we set the spin configuration obtained near the phase boundary as the initial spin configuration and perform the simulated annealing starting at low temperatures ($T_0=0.02$-$0.05$).
We identify magnetic phases by calculating the spin structure factor and spin scalar chirality in each sublattice (see Supplemental Material).

To judge whether the obtained spin configurations are topologically nontrivial, we compute a total skyrmion number ($N^\mathrm{tot}_\mathrm{sk}$) and a staggered skyrmion number ($N^\mathrm{stagg}_\mathrm{sk}$) as
\begin{align}
\label{eq:Ntot}
N^\mathrm{tot}_\mathrm{sk} &= |N^\mathrm{A}_\mathrm{sk}+N^\mathrm{B}_\mathrm{sk}|, \\ 
\label{eq:Nstagg}
N^\mathrm{stagg}_\mathrm{sk} &= |N^\mathrm{A}_\mathrm{sk}-N^\mathrm{B}_\mathrm{sk}|. 
\end{align}
By using them, we categorize a topological property into four types: ferrochiral (FC) SkX with $N^\mathrm{tot}_\mathrm{sk}\neq0$ and $N^\mathrm{stagg}_\mathrm{sk}=0$, antiferrochiral (AFC) SkX with $N^\mathrm{tot}_\mathrm{sk}=0$ and $N^\mathrm{stagg}_\mathrm{sk}\neq0$, ferrichiral (FerriC) SkX with $N^\mathrm{tot}_\mathrm{sk}\neq0$ and $N^\mathrm{stagg}_\mathrm{sk}\neq0$, and trivial states with $N^\mathrm{tot}_\mathrm{sk}=0$ and $N^\mathrm{stagg}_\mathrm{sk}=0$.

Figure~\ref{fig:Zero}(a) shows the phase diagram on the $\Theta$-$K$ ($|X^\mathrm{AB}|$-$K$) plane obtained by simulated annealing. 
By introducing $K$, the single-$Q$ spiral state shows the instabilities toward three multiple-$Q$ states: chiral stripe (CS) I, CS II, and FC (AFC) SkX II for positive (negative) $X^\mathrm{AB}$. 
The spin configurations of two CS states are characterized by a double-$Q$ superposition of the spiral wave at $\bm{Q}^{\rm spiral}_\alpha \equiv \bm{Q}_\eta$ and the sinusoidal wave at $\bm{Q}^{\rm sin}_\alpha \equiv \bm{Q}_{\eta'}$ ($\eta \neq \eta'$) in each sublattice $\alpha$~\cite{Ozawa_doi:10.7566/JPSJ.85.103703,Hayami_PhysRevB.95.224424,yambe2020double,doi:10.7566/JPSJ.91.093702}.
Although their sinusoidal component is the same for the sublattices A and B in both CS states, i.e., $\bm{Q}^{\rm sin}_{\mathrm{A}}=\bm{Q}^{\rm sin}_{\mathrm{B}}$, the spiral component is different from each other: $\bm{Q}^{\rm spiral}_{\mathrm{A}} \neq \bm{Q}^{\rm spiral}_{\mathrm{B}}$ for the CS I and $\bm{Q}^{\rm spiral}_{\mathrm{A}} = \bm{Q}^{\rm spiral}_{\mathrm{B}}$ for the CS II.
The CS states are the trivial states without $N^\mathrm{tot}_\mathrm{sk}$ and $N^\mathrm{stagg}_\mathrm{sk}$. 

Meanwhile, the FC (AFC) SkX II appears in $K \gtrsim 0.095$ and positive (negative) $X^\mathrm{AB}$, which indicates that a nonzero but small $\Theta \gtrsim 15.5\pi/240$ ($|X^\mathrm{AB}| \gtrsim 0.041$) is enough to stabilize the SkXs.
In both SkXs, the spin configurations on each sublattice are characterized by the triangular lattice of the skyrmion, while the constituent skyrmions are different, as shown in Figs.~\ref{fig:Zero}(b) and \ref{fig:Zero}(c): $(p^{\alpha},v^{\alpha},N^{\alpha}_\mathrm{sk})=(-1,+2,-2)$ for $\alpha={\rm A}, {\rm B}$ [Fig.~\ref{fig:skyrmion}(d)] in the FC SkX I and $(p^{\mathrm{A}},v^{\mathrm{A}},N^{\mathrm{A}}_\mathrm{sk})=(+1,+2,+2)$ [Fig.~\ref{fig:skyrmion}(e)] and $(p^{\mathrm{B}},v^{\mathrm{B}},N^{\mathrm{B}}_\mathrm{sk})=(-1,+2,-2)$ [Fig.~\ref{fig:skyrmion}(d)] in the AFC SkX II.
In the end, the FC SkX II has the uniform skyrmion number as $N^\mathrm{tot}_\mathrm{sk}=4$ and $N^\mathrm{stagg}_\mathrm{sk}=0$, whereas the AFC SkX II has the staggered skrymion number as  $N^\mathrm{tot}_\mathrm{sk}=0$ and $N^\mathrm{stagg}_\mathrm{sk}=4$. 
The key ingredients for the FC/AFC SkX II are the (anti)ferromagnetically coupled bipartite structure and itinerant nature giving nonzero positive $K$.  
We find that the AFC SkX II is regarded as the antiferromagnetic SkX since the opposite sign on the skyrmion number arises from the opposite polarity. 
Note that such an antiferromagnetic SkX with a high topological number has not been found so far.
Furthermore, our mechanism does not require a multi-layer structure to stabilize the antiferromagnetic SkX~\cite{Gobel_PhysRevB.96.060406,PhysRevB.105.075102}. 　

\begin{figure*}[t!]
\begin{center}
\includegraphics[width=1.0\hsize]{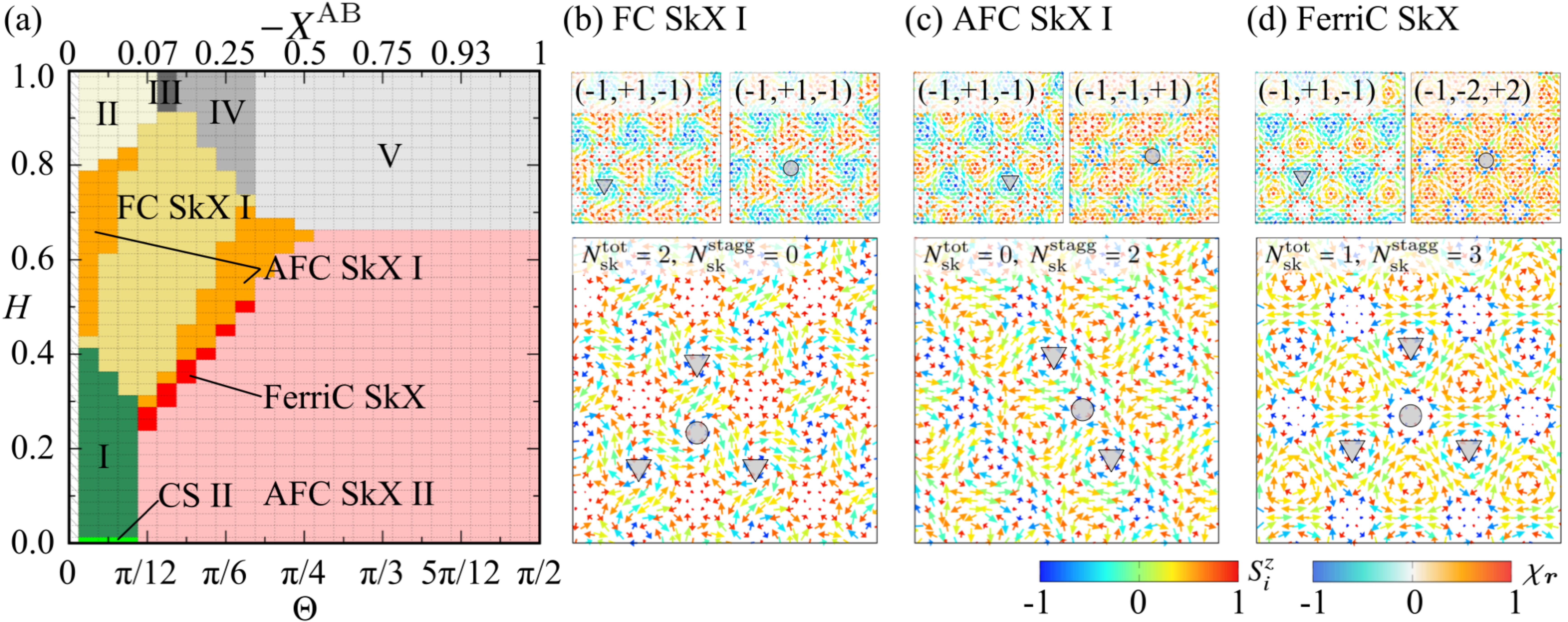} 
\caption{
\label{fig:Field}
(a) Phase diagram on the $\Theta$-$H$ plane at $X^\mathrm{AB}<0$, $K=0.4$, and $T=0.01$.
FC SkX, AFC SkX, and FerriC SkX represent the ferrochiral SkX, antiferrochiral SkX, and ferrichiral SkX, respectively.
I-V are nontopological phases. 
Snapshots of (b) the FC SkX I at $\Theta=\pi/8$ and $H=0.5$, (c) AFC SkX I at $\Theta=\pi/8$ and $H=0.4$, and (d) FerriC SkX at $\Theta=\pi/8$ and $H=0.375$, which corresponds to Figs.~\ref{fig:Zero}(b) and \ref{fig:Zero}(c).
}
\end{center}
\end{figure*}

We further show rich topological spin textures by introducing the magnetic-field term $\mathcal{H}^\mathrm{Zeeman}=-H\sum_{\alpha,i}S_{\alpha i}^z$, where we set $H>0$ favoring the negative polarity of the skyrmion.
Figure~\ref{fig:Field}(a) shows the $\Theta$-$H$ phase diagram at fixed $X^\mathrm{AB}<0$ and $K=0.4$. 
We find multiple SkX instabilities driven by the magnetic field for $\Theta\le\pi/4$ ($|X^\mathrm{AB}|\le 0.5$): 
FC SkX I, AFC SkX I, and FerriC SkX.
We discuss the details of the spin configurations in each field-induced SkX in the following. 
We detail the other trivial phases denoted as I-V in Supplemental Material. 

The FC SkX I is constituted of the skyrmion with  $(p^{\alpha},v^{\alpha},N^{\alpha}_\mathrm{sk})=(-1,+1,-1)$ [Fig.~\ref{fig:skyrmion}(a)] for both sublattices ($\alpha=$A, B), as shown in Fig.~\ref{fig:Field}(b), whose vorticity and the skyrmion number is halved compared to the FC SkX II in Fig.~\ref{fig:Zero}(b), i.e., $N^\mathrm{tot}_\mathrm{sk}=2$ and $N^\mathrm{stagg}_\mathrm{sk}=0$. 
Meanwhile, the FC SkX I is stabilized even for negative $X^\mathrm{AB}$, which is qualitatively in contrast to the FC SkX II stabilized only for positive $X^\mathrm{AB}$. 
The emergence of the FC SkX I is due to the different skyrmion core positions to gain the energy by $J X^{\mathrm{AB}}$; the skyrmion core at the B sublattice represented by the circle is separated from that at the A sublattice by the triangle so as to form the honeycomb network, as shown in Fig.~\ref{fig:Field}(b).
A similar shift of core positions has been found in three-sublattice SkXs in the antiferromagnetic triangular and kagome lattices~\cite{Rosales_PhysRevB.92.214439,PhysRevB.100.245106}. 

The AFC SkX I consists of the skyrmions with $(p^{\mathrm{A}},v^{\mathrm{A}},N^{\mathrm{A}}_\mathrm{sk})=(-1,+1,-1)$ [Fig.~\ref{fig:skyrmion}(a)] and $(p^{\mathrm{B}},v^{\mathrm{B}},N^{\mathrm{B}}_\mathrm{sk})=(-1,-1,+1)$ [Fig.~\ref{fig:skyrmion}(c)], as shown in Fig.~\ref{fig:Field}(c), which results in $N^\mathrm{tot}_\mathrm{sk}=0$ and $N^\mathrm{stagg}_\mathrm{sk}=2$.
In contrast to the AFC SkX II, the opposite sign of $N^{\alpha}_\mathrm{sk}$ in the AFC SkX I is owing to the opposite vorticity in the constituent skyrmions instead of the polarity; the A-sublattice SkX is characterized by the skyrmion spin texture with $v^{\mathrm{A}}=+1$, while the B-sublattice one is described by the anti-skyrmion spin texture with $v^{\mathrm{B}}=-1$ in the upper panels of Fig.~\ref{fig:Field}(c). 
Thus, the AFC SkX I is regarded as a coexisting state of the skyrmion and anti-skyrmion, where they are degenerate in the present model without the spin-orbit coupling.
Although the domain structure of the skyrmion and anti-skyrmion has been found~\cite{Okubo_PhysRevLett.108.017206,Mitsumoto_PhysRevB.105.094427}, their coexisting ordered state has not been discovered without the multi-layer structure so far~\cite{Hayami_PhysRevB.105.184426}.
The (anti-)skyrmion cores in the AFC SkX I are aligned in a one-dimensional way breaking the threefold rotational symmetry.

The FerriC SkX sandwiched by the AFC SkX II and AFC SkX I is formed by the skyrmions with $(p^{\mathrm{A}},v^{\mathrm{A}},N^{\mathrm{A}}_\mathrm{sk})=(-1,+1,-1)$ [Fig.~\ref{fig:skyrmion}(a)] and $(p^{\mathrm{B}},v^{\mathrm{B}},N^{\mathrm{B}}_\mathrm{sk})=(-1,-2,+2)$ [Fig.~\ref{fig:skyrmion}(f)], as shown in Fig.~\ref{fig:Field}(d). 
Thus, this state has nonzero $N^\mathrm{tot}_\mathrm{sk}=1$ and $N^\mathrm{stagg}_\mathrm{sk}=3$ with a ferri-type alignment of $N^\alpha_\mathrm{sk}$. 
The appearance of the FerriC SkX might be owing to the multi-sublattice system in itinerant magnets since the SkX with $|N^{\alpha}_\mathrm{sk}|=2$ is only realized for nonzero $K$. 
Indeed, such instability has not been reported in the isotropic localized spin model. 
The FerriC SkX is quite different from the other sublattice SkXs reported in the present study and previous studies: In the previous findings, the spin configurations on each sublattice are usually energetically degenerate, while the FerriC SkX consists of two SkXs with different energy.
The stabilization of this novel intermediate state between the AFC SkXs I and II is a consequence of the competition between the effective spin interactions and Zeeman effect: The former tends to favor the SkX with $|N_\mathrm{sk}|=2$, while the latter tends to favor the SkX with $|N_\mathrm{sk}|=1$~\cite{Ozawa_PhysRevLett.118.147205,yambe2021skyrmion}.

To summarize, we theoretically propose a rich variety of SkXs in the itinerant honeycomb magnet by the synergy of the spin, charge, and sublattice degrees of freedom.
By constructing the ground-state phase diagram of the effective spin model for the KLM, we find five new topological spin textures with different topological properties: ferrochiral SkXs I and II, antiferrochiral SkXs I and II, and ferrichiral SkX.
We demonstrate that the essence lies in the competition among the bilinear interaction between different sublattices, positive biquadratic interaction, and magnetic field. 
Since the total (staggered) skyrmion number is closely related to the emergence of the topological Hall (spin Hall) effect, one expects a variety of transport phenomena driven by the emergent electromagnetic field.
The important conditions to induce the present multiple SkX instabilities are the positive biquadratic interaction arising from the itinerant nature, bipartite lattice structure, and the negligibly small spin-orbit coupling. 

\begin{acknowledgments}
This research was supported by JSPS KAKENHI Grants Numbers JP21H01037, JP22H04468, JP22H00101, JP22H01183, and by JST PRESTO (JPMJPR20L8). 
R.Y. was supported by Forefront Physics and Mathematics Program to Drive Transformation (FoPM).
Parts of the numerical calculations were performed in the supercomputing systems in ISSP, the University of Tokyo.
\end{acknowledgments}

\bibliography{ms.bbl}
\end{document}


\title{Supplemental material for "Ferrochiral, antiferrochiral, and ferrichiral skyrmion crystals in an itinerant  honeycomb magnet"}

\author{Ryota Yambe}
\affiliation{Department of Applied Physics, The University of Tokyo, Tokyo 113-8656, Japan }
\author{Satoru Hayami}
\affiliation{Graduate School of Science, Hokkaido University, Sapporo 060-0810, Japan}

\maketitle

\section{Spin- and chirality-related quantities}

In the simulation, we identify magnetic phases from the spin structure factor, magnetization, spin scalar chirality, and skyrmion number in each sublattice.
The spin structure factor for the sublattice $\alpha=\mathrm{A},\mathrm{B}$ is defined as 
\begin{align}
S^\mu(\alpha,\bm{q}) = \left\langle\frac{1}{N}\sum_{j,k}S^\mu_{\alpha j}S^\mu_{\alpha k} e^{i\bm{q}\cdot(\bm{R}_{\alpha j}-\bm{R}_{\alpha k})}\right\rangle,
\end{align}
where $\mu=x,y,z$, $\bm{R}_{\alpha j}$ is the position vector at site $j$ on the sublattice $\alpha$, $N$ is the number of the unit cell, and $\langle\cdots\rangle$ is the average over the Monte Carlo samples.
We also calculate the in-plane spin structure factor, $S^\perp(\alpha,\bm{q})=S^x(\alpha,\bm{q})+S^y(\alpha,\bm{q})$.
The magnetization for the sablattice $\alpha$ is defined as 
\begin{align}
M_\alpha = \left\langle\frac{1}{N}\sum_{j}S^z_{\alpha j}\right\rangle.
\end{align}

The local spin scalar chirality of the triangle on the sublattice $\alpha$ is defined as
\begin{align}
 \chi_{\alpha \bm{r}}= \bm{S}_{\alpha j}\cdot( \bm{S}_{\alpha k}\times \bm{S}_{\alpha l}),
 \end{align}
 where the position vector $\bm{r}$ represents the triangle center and the triangle consists of $(j,k,l)$ sites labeled in the counterclockwise order.
The uniform spin scalar chirality of the sublattice $\alpha$ is given by 
\begin{align}
\chi_\alpha^\mathrm{sc} =  \left\langle \frac{1}{N}\sum_{\bm{r}} \chi_{\alpha\bm{r}} \right\rangle.
\end{align}

The skyrmion density at the triangle $\bm{r}$ on the sublattice $\alpha$~\cite{BERG1981412} is defined as  
\begin{align}
\tan\left(\frac{\Omega_{\alpha \bm{r}}}{2}\right) = \left[\frac{\bm{S}_{\alpha j}\cdot( \bm{S}_{\alpha k}\times \bm{S}_{\alpha l})}{1+\bm{S}_{\alpha j}\cdot\bm{S}_{\alpha k}+\bm{S}_{\alpha k}\cdot\bm{S}_{\alpha l}+\bm{S}_{\alpha l}\cdot\bm{S}_{\alpha j}} \right].
\end{align}
Then, the skyrmion number for the sublattice $\alpha$ is given by  
\begin{align}
\label{eq:Nsk}
N^\alpha_\mathrm{sk} = \frac{1}{4\pi N_\mathrm{m}}\left\langle \sum_{\bm{r}} \Omega_{\alpha \bm{r}} \right\rangle,
\end{align}
where $N_\mathrm{m}$ is the number of the magnetic unit cell.

\section{Trivial phases I-V}

\begin{figure*}[h!]
\begin{center}
\includegraphics[width=1.0\hsize]{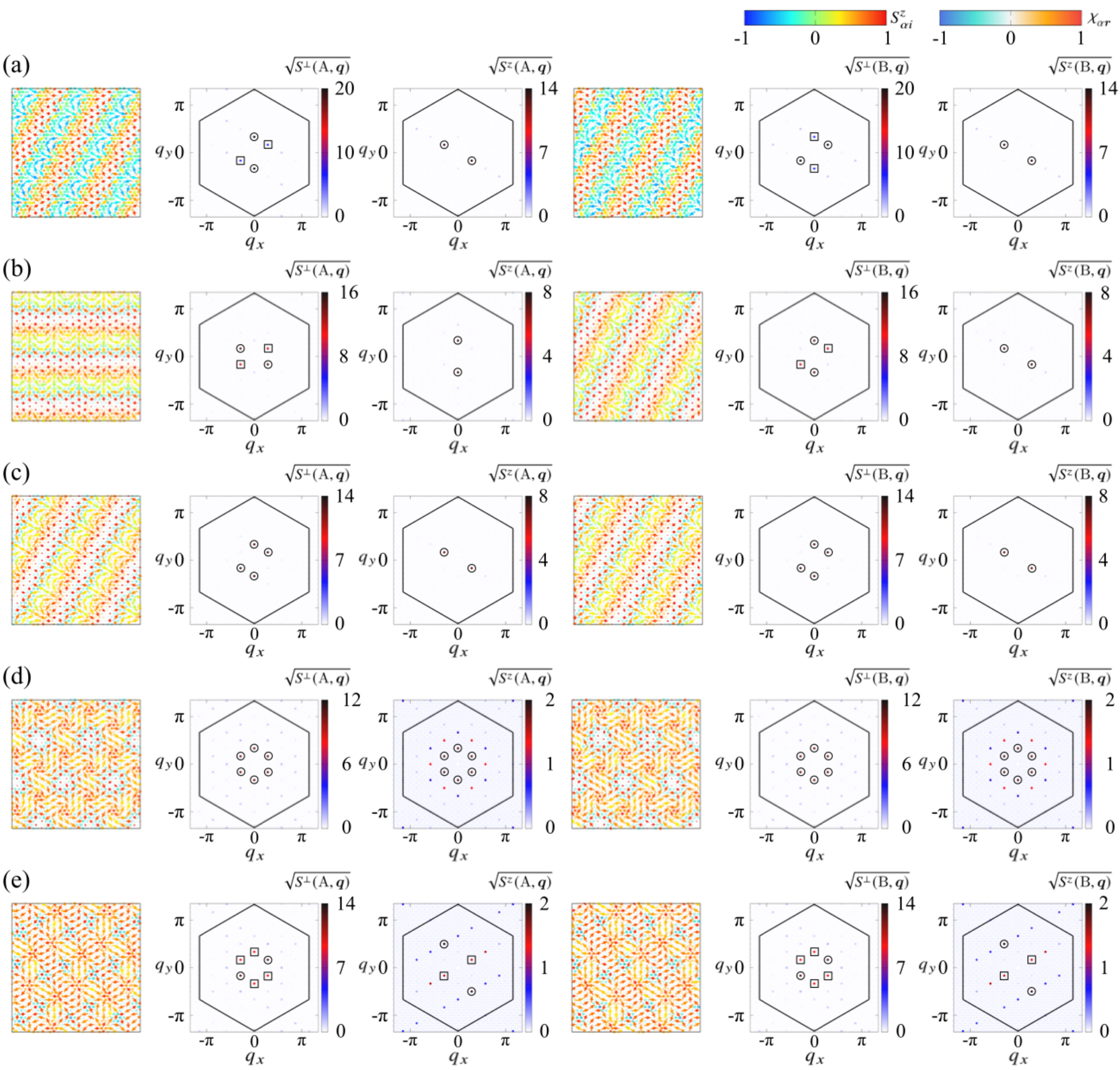} 
\caption{
\label{fig:3q}
First (fourth) column: Spin ($\bm{S}_{\alpha i}$) and chirality  ($\chi_{\alpha \bm{r}}$) configurations on the sublattice A (B) of
(a) the trivial state I at $\Theta=\pi/24$ and $H=0.3$,
(b) the trivial state II at $\Theta=\pi/24$ and $H=1$,
(c) the trivial state III at $\Theta=5\pi/48$ and $H=1$,
(d) the trivial state IV at $\Theta=\pi/6$ and $H=1$,
and (e) the trivial state V at $\Theta=\pi/4$ and $H=1$.
The arrows, contours of arrows, and contours of circles show the $xy$ spin component, $z$ spin component, and spin scalar chirality, respectively.  
Second and third (fifth and sixth) columns: The in-plane [$S^\perp(\alpha,\bm{q})$] and out-of-plane [$S^z(\alpha,\bm{q})$] spin structure factors for the sublattice A (B) in momentum space, respectively.
The circle and square highlight the dominant and subdominant peaks, respectively.
The hexagons with a solid line show the first Brillouin zone.
The $\bm{q}=\bm{0}$ component is removed for better visibility.
}
\end{center}
\end{figure*}

In Fig.~3, we find the field-induced trivial phases denoted as I-V with $N^\mathrm{tot}_\mathrm{sk}=N^\mathrm{stagg}_\mathrm{sk}=0$.
We show their spin configurations, chirality configurations, and spin structure factors in Fig.~\ref{fig:3q}.
Since all the phases exhibit the triple-$Q$ peaks at $\bm{Q}_1=(0,\pi/3)$, $\bm{Q}_2=(-\sqrt{3}\pi/6,-\pi/6)$, and $\bm{Q}_3=(\sqrt{3}\pi/6,-\pi/6)$ in the spin structure factor, they are regarded as the triple-$Q$ states. 
The trivial spin configurations I and II are characterized by the double-$Q$ in-plane spin structure factor and the single-$Q$ out-of-plane one, as shown in Figs.~\ref{fig:3q}(a) and \ref{fig:3q}(b), respectively, where the peak positions depend on the sublattices.
The trivial state I is characterized by the sublattice-dependent in-plane spin structure factor and sublattice-independent our-of-plane one: $S^\perp({\rm A},\bm{Q}_1)>S^\perp({\rm A},\bm{Q}_2)$ and $S^z({\rm A},\bm{Q}_3)$ for the sublattice A and $S^\perp({\rm B},\bm{Q}_2)>S^\perp({\rm B},\bm{Q}_1)$ and $S^z({\rm B},\bm{Q}_3)$ for the sublattice B.
The trivial phase II shows the sublattice-dependent in-plane and out-of-plane spin structure factors: $S^\perp({\rm A},\bm{Q}_3)>S^\perp({\rm A},\bm{Q}_2)$ and $S^z({\rm A},\bm{Q}_1)$ for the sublattice A and $S^\perp({\rm B},\bm{Q}_1)>S^\perp({\rm B},\bm{Q}_2)$ and $S^z({\rm B},\bm{Q}_3)$ for the sublattice B.
Meanwhile, the trivial spin configurations III-V show the sublattice-independent in-plane and out-of-plane spin structure factors: 
$S^\perp(\alpha,\bm{Q}_1)=S^\perp(\alpha,\bm{Q}_2)$ and $S^z(\alpha,\bm{Q}_3)$ for $\alpha=\mathrm{A},\mathrm{B}$ in the phase III,
$S^{\perp(z)}(\alpha,\bm{Q}_1)=S^{\perp(z)}(\alpha,\bm{Q}_2)=S^{\perp(z)}(\alpha,\bm{Q}_3)$ for $\alpha=\mathrm{A},\mathrm{B}$ in the phase IV, 
and $S^\perp(\alpha,\bm{Q}_2)>S^\perp(\alpha,\bm{Q}_1)=S^\perp(\alpha,\bm{Q}_3)$ and $S^z(\alpha,\bm{Q}_1-\bm{Q}_3)>S^z(\alpha,\bm{Q}_2)$ for $\alpha=\mathrm{A},\mathrm{B}$ in the phase V.

\section{Skyrmion crystals in the models with different form factors}

We consider the effective spin model with different form factors for the bilinear and biquadratic interactions, which is given by
\begin{align}
\mathcal{H}^\mathrm{eff}&=-2J\sum_{\eta}\sum_{\alpha,\beta}X^{\alpha\beta}\bm{S}_{\alpha \bm{Q}_\eta}\cdot\bm{S}_{\beta -\bm{Q}_\eta} \nonumber\\
&+2\frac{K}{N}\sum_{\eta}\left(\sum_{\alpha,\beta}X^{\alpha\beta}_\mathrm{K}\bm{S}_{\alpha \bm{Q}_\eta}\cdot\bm{S}_{\beta -\bm{Q}_\eta}\right)^2 \nonumber \\
&-H\sum_{\alpha,i}S_{\alpha i}^z.
\end{align} 
Here,  $X^{\alpha\beta}$ and $X^{\alpha\beta}_\mathrm{K}$ are the form factors for the RKKY interaction and biquadratic interaction, respectively.
Although we consider the case of $X^{\alpha\beta}=X^{\alpha\beta}_\mathrm{K}$ in the main text for simplicity, the form factors are usually different. 
In this section, we show that the SkX phases in Fig.~3 also apper even for $X^{\alpha\beta}\neq X^{\alpha\beta}_\mathrm{K}$.
Specifically, we consider three different $X^{\alpha\beta}_\mathrm{K}$ while fixing $X^{\rm AA}\equiv \cos^2\Theta$ and $X^{\rm AB}\equiv -\sin^2\Theta$ to cover various situations:
(i) $X^{\rm AA}_\mathrm{K}\equiv \cos^2\Theta$ and $X^{\rm AB}_\mathrm{K}\equiv \sin^2\Theta$,
(ii) $X^{\rm AA}_\mathrm{K}\equiv \sin^2\Theta$ and $X^{\rm AB}_\mathrm{K}\equiv -\cos^2\Theta$,
and (iii) $X^{\rm AA}_\mathrm{K}\equiv \sin^2\Theta$ and $X^{\rm AB}_\mathrm{K}\equiv \cos^2\Theta$.
By following the same manner of the simulation in the main text, we find the FC SkX I, AFC SkXs I and II, and FerriC SkX in various $\Theta$ and $H$.
For example, we show critical magnetic fields for multiple topological transitions in Table~\ref{tab:Hc}, where the AFC SkX II appears at low fields including $H=0$, the FerriC SkX appears above $H_{{\rm c} 1}$, the AFC SkX I appears above $H_{{\rm c} 2}$, the FC SkX I appears above $H_{{\rm c} 3}$, and other trivial states appear above $H_{{\rm c} 4}$ while increasing $H$.
In this way, a variety of the SkXs discussed in the main text are ubiquitously stabilized in the effective spin model with the antiferromagnetic intersublattice RKKY interaction and the positive biquadratic interaction irrespective of the form factor $X^{\alpha\beta}_\mathrm{K}$.

\begin{table}[h!]
\caption{\label{tab:Hc}
Critical magnetic fields $H_{{\rm c} 1}$, $H_{{\rm c} 2}$, $H_{{\rm c} 3}$ and $H_{{\rm c} 4}$ for multiple topological transitions at specific $\Theta$.
Here, we find the AFC SkX II at $0\le H < H_{{\rm c} 1}$, the FerriC SkX at $ H_{{\rm c} 1}< H < H_{{\rm c} 2}$, the AFC SkX I at $ H_{{\rm c} 2}< H < H_{{\rm c} 3}$, and the FC SkX at $ H_{{\rm c} 3}< H < H_{{\rm c} 4}$.
}
\begin{ruledtabular}
\begin{tabular}{llllll}
  & $\Theta$ & $H_{{\rm c}1}$ & $H_{{\rm c}2}$ & $H_{{\rm c}3}$ & $H_{{\rm c}4}$ \\ \hline
case (i) & $5\pi /48$ & 0.3875 & 0.4125 & 0.4825 & 0.8625 \\ 
case (ii) & $\pi /6$ & 0.2625 & 0.3375 & 0.4625 & 0.9125\\
case (iii) & $7\pi /48$ & 0.3375 & 0.3625 & 0.5375 & 0.9625 
\end{tabular}
\end{ruledtabular}
\end{table}

\bibliography{supplement.bbl}